\documentclass[a4paper]{jpconf}
\usepackage{graphicx}
\usepackage{hyperref}

\begin{document}
\title{Einstein-Maxwell-Chern-Simons
Black Holes}

\author{J Kunz$^1$, J L Bl\'azquez-Salcedo$^1$, 
F Navarro-L\'erida$^2$, E Radu$^3$}

\address{$^1$ Institut f\"ur  Physik, Universit\"at Oldenburg,
D-26111 Oldenburg, Germany}

\address{$^2$ Dept.~de F\'{\i}sica At\'omica, Molecular y Nuclear, 
Ciencias F\'{\i}sicas,
Universidad Complutense de Madrid, E-28040 Madrid, Spain}

\address{$^3$ Departamento de F\'isica da Universidade de Aveiro and CIDMA,
 Campus de Santiago, 3810-183 Aveiro, Portugal}

\ead{jutta.kunz@uol.de,
jose.blazquez.salcedo@uol.de,
fnavarro@fis.ucm.es,
eugen.radu@ua.pt}

\begin{abstract}
Black holes in 5-dimensional Einstein-Maxwell-Chern-Simons (EMCS) theory 
and their intriguing properties are discussed. 
For the special case of the CS coupling constant $\lambda=\lambda_{SG}$,
as obtained from supergravity, a closed form solution is known 
for the rotating black holes. 
Beyond this supergravity value, the EMCS black hole solutions can 
e.g.~exhibit nonuniqueness and form sequences of radially excited solutions. 
In the presence of a negative cosmological constant the
black holes can possess an extra-parameter corresponding 
to a magnetic flux in addition to the mass, electric charge 
and angular momenta. 
This latter family of black holes possesses also a solitonic limit. 
Finally, a new class of squashed EMCS black hole solutions is discussed.
\end{abstract}

\section{Introduction}

The properties of black holes in higher dimensions can differ in many
respects from those known in four dimensions. 
Most prominently, asymptotically flat vacuum black holes may possess
a non-spherical horizon topology in more than four dimensions.
But also for black holes with a spherical horizon suprises arise,
as soon as a U(1) gauge field is coupled, i.e., Einstein-Maxwell (EM)
black holes are considered. 
In odd dimensions the presence of a Chern-Simons (CS) term 
allows for further intriguing properties of the resulting 
Einstein-Maxwell-Chern-Simons (EMCS) black holes.
In the following first asymptotically flat black hole solutions of EMCS theory
will be discussed. Then a negative cosmological constant will be included,
making contact with the celebrated AdS/CFT correspondence,
a central topic of the conference.

\section{Asymptotically Flat EMCS Black Holes}

In odd dimensions $D=2n+1$ the Einstein-Maxwell action may be
supplemented by an `${\cal A} {\cal F}^n$' Chern-Simons term.
In 5 dimensions the EMCS action reads
\begin{equation}
S= \int \frac{1}{16\pi G_5} \Big\{
\sqrt{-g} \left( R - {\cal F}_{mn} {\cal F}^{mn} \right)
 -  \frac{2  \lambda}{3\sqrt{3}}
\varepsilon^{mnpqr} 
{\cal A}_m {\cal F}_{np} {\cal F}_{qr} 
\Big\} d^5x  
\end{equation}
with Newton's constant $G_5$, curvature scalar $R$, 
gauge potential ${\cal A}_m$, field strength tensor ${\cal F}_{mn}$,
and CS coupling constant ${\lambda}$.
In EM theory $\lambda=0$, while $\lambda=1$ in the bosonic sector
of minimal $D=5$ supergravity.
Variation of the action leads to the Einstein equations ($8\pi G=1$)
\begin{equation}
G_{mn}=2 \left( {\cal F}_{mr} {\cal F}^r_{\ n} 
  - \frac{1}{4} {\cal F}_{rs} {\cal F}^{rs} \right) ,
\end{equation}
which are unchanged w.r.t.~the pure Maxwell case,
and to the Maxwell-CS equations
\begin{equation}
\nabla_n {\cal F}^{mn} + \frac{\lambda}{2\sqrt{3}}
\epsilon^{mnpqr} {\cal F}_{np} {\cal F}_{qr} =0 .
\end{equation}
Clearly, the CS term breaks the charge symmetry $Q\to - Q$
present in Maxwell theory.

\subsection{$D=5$ Einstein-Maxwell Theory}

The $D$-dimensional generalizations of the Kerr black holes
are given by the Myers-Perry (MP) black holes,
which possess $N=\lfloor \frac{D-1}{2} \rfloor$ independent
angular momenta $J_i$, $i=1,\dots,N$
\cite{Myers:1986un}.
The inclusion of a Maxwell field, however, 
most of the time prevents the construction
of black hole solutions in closed form,
even if in odd dimensions
the angular momenta are chosen to be of equal magnitude,
i.e., when only cohomogeneity-1 solutions are sought.
In five dimensions an appropriate ansatz
for the metric and the gauge potential is given by
\begin{eqnarray}
&&ds^2 = -F_0(r) dt^2 + F_1(r)dr^2
  + \frac{1}{4}F_2(r)(\sigma_1^2+\sigma_2^2)
+\frac{1}{4}F_3(r) \big(\sigma_3-2\omega(r) dt \big)^2 , 
\label{at0} \\
&&{\cal A}=a_0(r)dt +a_\varphi(r) \frac{1}{2} \sigma_3
\label{at1}
\end{eqnarray}
with $\sigma_1^2+\sigma_2^2=d\bar \theta^2+\sin^2\bar
\theta d\psi^2$, $\sigma_3=d\phi+\cos \bar \theta d \psi$, 
and functions $F_0,\dots,F_3$, $\omega$, $a_0$ and $a_\varphi$.

In this case (i) perturbative calculations in the charge
or in the rotation parameter can be performed,
(ii) the solutions can be obtained numerically, or
(iii) near-horizon solutions can be found in the extremal limit.
To obtain the latter, an adequate ansatz for the metric 
and the gauge potential in five dimensions is given by
\cite{Kunduri:2013gce}
\begin{eqnarray}
&& ds^2=v_1(\frac{dr^2}{r^2}-r^2 dt^2) 
+ v_2 \left[\sigma_1^2+\sigma_2^2
+ v_3(\sigma_3- k r dt)^2 \right] , \\
&& 
{\cal A} =
 q_1 r dt
+ q_2 \left( \sigma_3 - k r dt \right)
\label{at2}
\end{eqnarray}
with constants $v_1$, $v_2$, $v_3$, $q_1$, $q_2$ and $k$.

\begin{figure}[h]
\begin{minipage}{17pc}
\hspace{-1pc}
\includegraphics[width=14pc,angle=270]{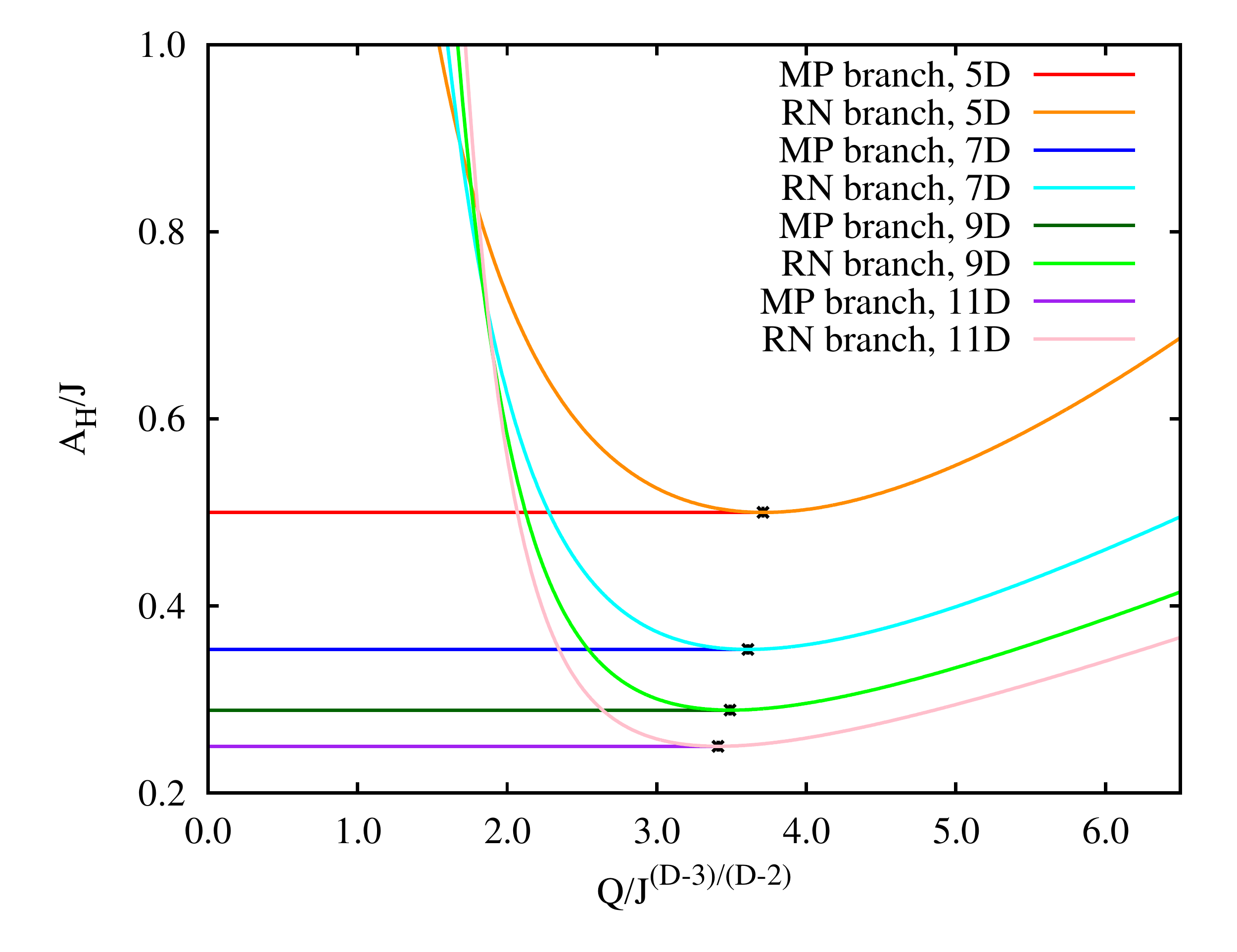}
\caption{\label{NH1}
Odd-$D$ EM near-horizon solutions:
area vs charge (scaled with appropriate powers of the angular momentum).}
\end{minipage}
\hspace{4pc}%
\begin{minipage}{17pc}
\hspace{-2pc}
\includegraphics[width=14pc,angle=270]{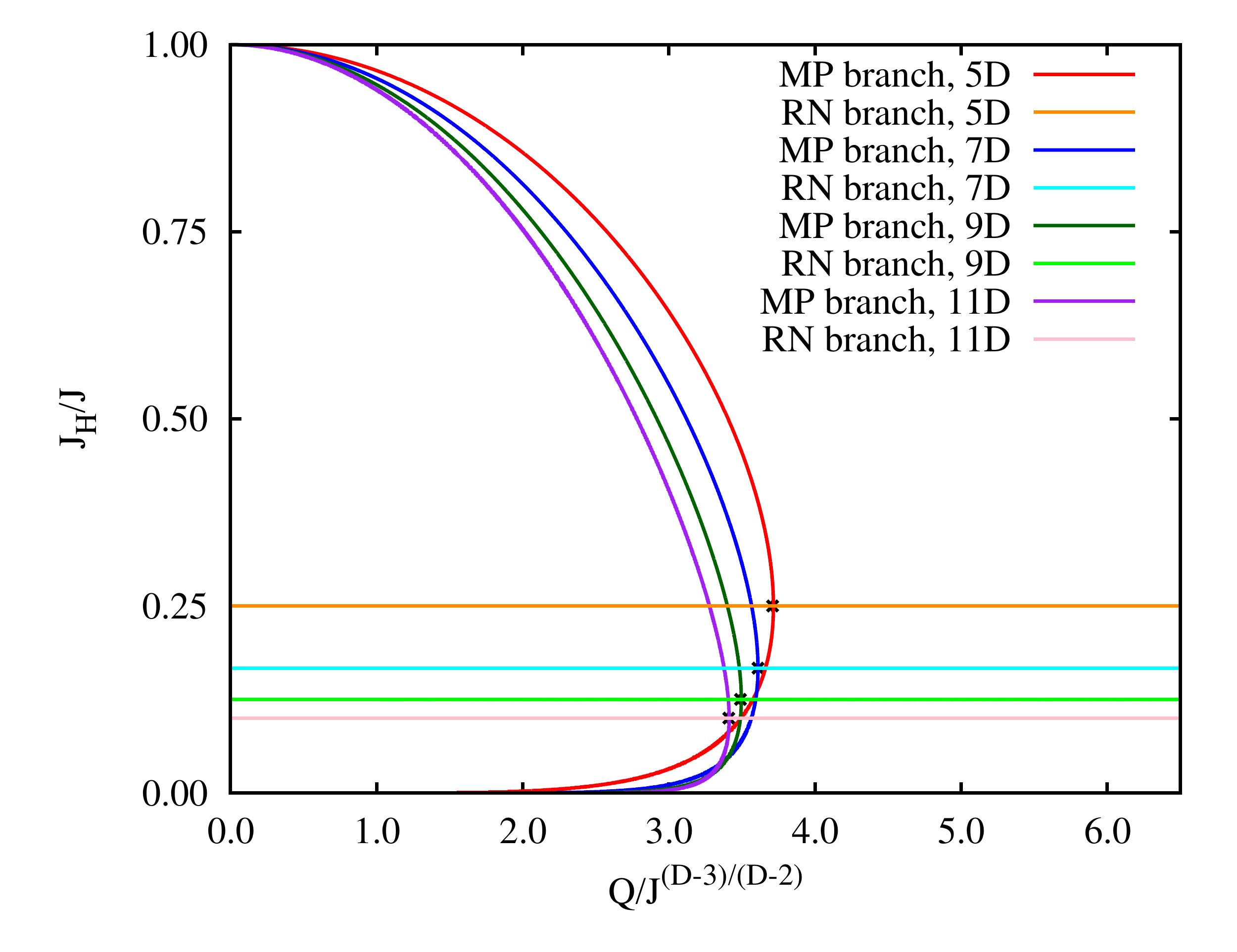}
\caption{\label{NH2}
Same as Fig.~\ref{NH1}
for the horizon angular momentum.
}
\end{minipage}
\end{figure}

An analogous ansatz holds in more than five dimensions.
Solving the resulting algebraic system of equations for the constants leads to
two distinct solutions.
The first solution starts at the extremal MP black hole
and satisfies $J= \sqrt{2(D-3)} A_{\rm H}$, thus its angular momentum
is proportional to its horizon area.
The second solution starts at the extremal Reissner-Nordstr\"om (RN) black hole
and satisfies $J= (D-1) J_{\rm H} $, thus its angular momentum
is proportional to its horizon angular momentum.
As seen in Figs.~\ref{NH1},\ref{NH2}
both branches intersect at a critical solution 
(black cross) and continue beyond.
Interestingly, only the name-giving parts of the MP and RN solutions
up to the intersection point are realized globally
\cite{Blazquez-Salcedo:2013yba,Blazquez-Salcedo:2013wka}.
The domain of existence of these EM black holes is seen in
Fig.~\ref{DE1} and corresponds to the area enclosed by the $\lambda=0$ curve.

\subsection{$D=5$ minimal supergravity}

\begin{figure}[h]
\begin{minipage}{17pc}
\includegraphics[width=18pc,angle=0]{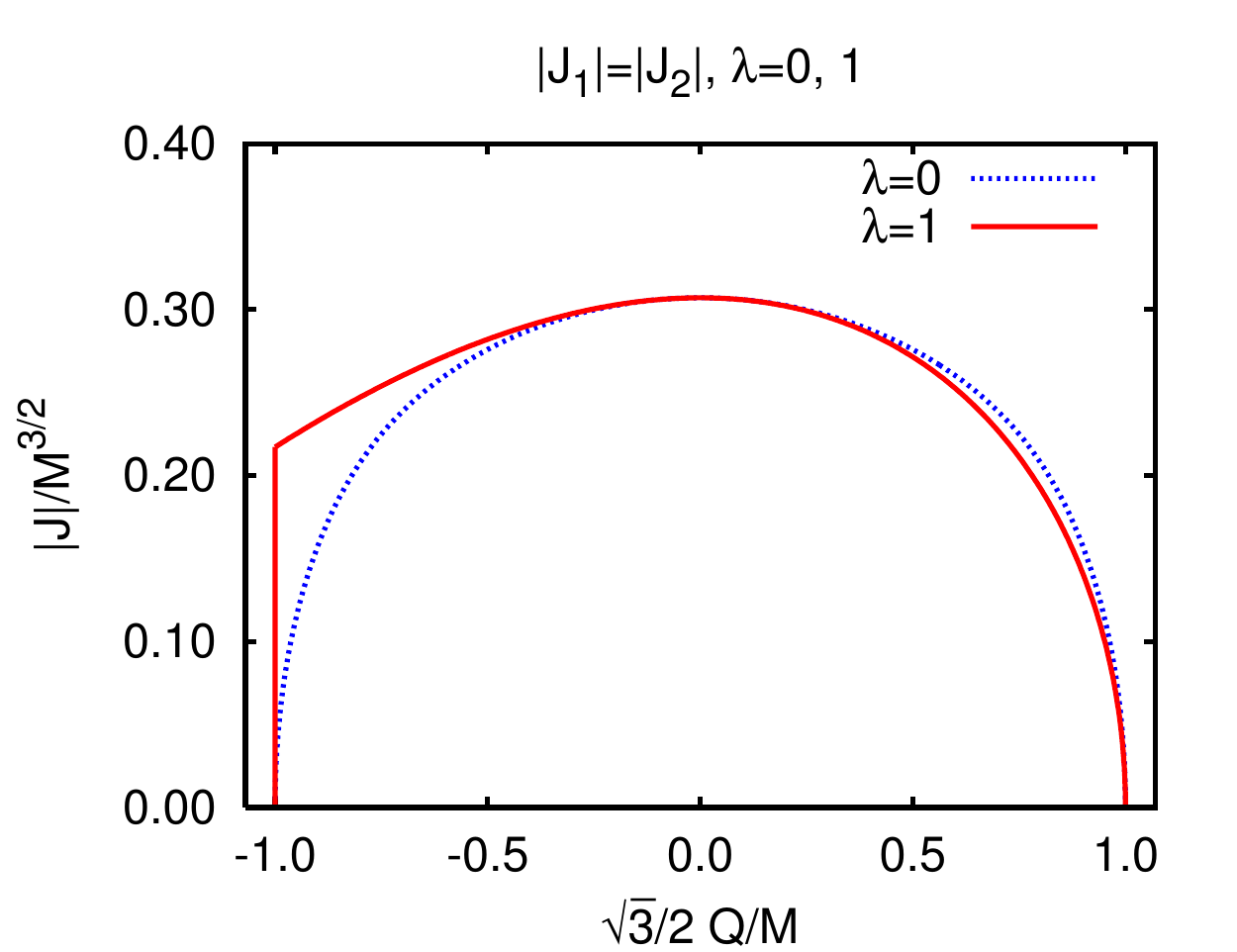}
\caption{\label{DE1}
Domain of existence of $D=5$ EMCS black holes:
angular momentum vs charge (scaled with appropriate powers of the mass)
for CS coupling $\lambda=1$ and $\lambda=0$.
}
\end{minipage}
\hspace{4pc}%
\begin{minipage}{17pc}
\includegraphics[width=18pc,angle=0]{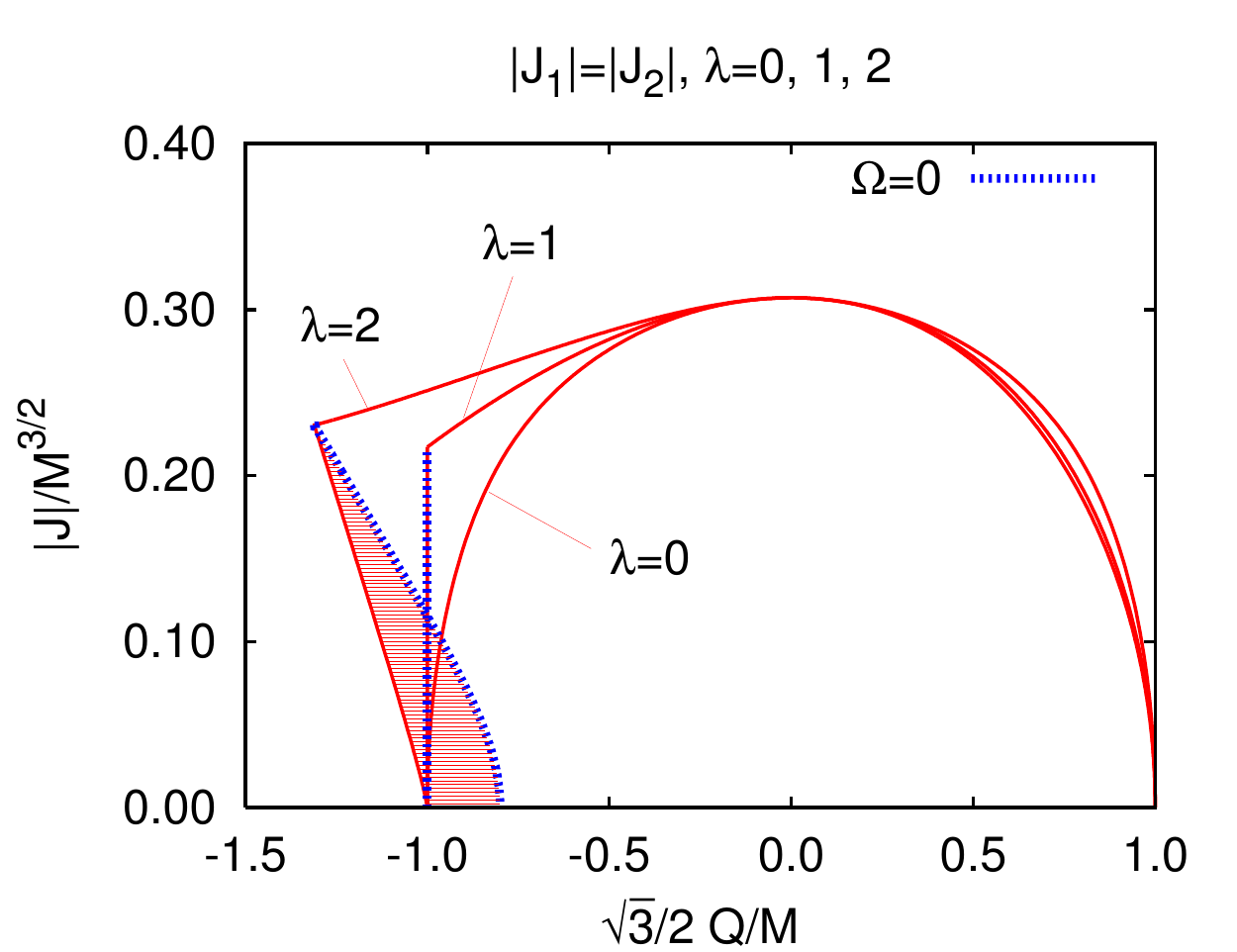}
\caption{\label{DE2}
Same as Fig.~\ref{DE1} including CS coupling $\lambda=2$.
The shaded area represents counterrotating black holes.
$\Omega=0$ black holes are also indicated.
}
\end{minipage}
\end{figure}

For the case $\lambda=1$ representing $D=5$ minimal supergravity
the black hole solutions are known in closed form
\cite{Breckenridge:1996is,Cvetic:2004hs,Chong:2005hr}.
The domain of existence is also seen in Fig.~\ref{DE1}
and delimited by the $\lambda=1$ curve.
The charge symmetry $Q\to -Q$ is clearly broken here.
The vertical line represents the extremal BMPV black holes 
\cite{Breckenridge:1996is}.
For these black holes, as the charge $Q$ is kept fixed and 
the angular momentum $J$ increases, the mass $M$ remains constant.

When considering the first law 
$dM = T dS +  2 \Omega dJ + \Phi dQ$
for these BMPV black holes,
one realizes that this implies
that their horizon angular velocity $\Omega$
must be zero, while the global angular momentum is finite.
Thus angular momentum is stored in the Maxwell field,
where a negative fraction of the angular momentum resides
behind the horizon. While
the effect of rotation is to deform the horizon into a squashed 3-sphere
\cite{Gauntlett:1998fz}.
The set of BMPV solutions ends at a critical solution with vanishing
area.

\subsection{$D=5$ EMCS theory: $\lambda \ne 1$}

Let us now consider the CS coupling as a free parameter
and increase it above the supergravity value \cite{Kunz:2005ei}.
Then it becomes clear that the supergravity value represents
the borderline between stability and instability,
since at $\lambda=1$ a zero mode is present,
leading to a rotational instability for larger values of $\lambda=1$
\cite{Gauntlett:1998fz,Kunz:2005ei}.

Then solutions with vanishing horizon angular velocity no longer
form a part of the boundary of the domain of existence
(except for the critical solution with vanishing area at the cusp),
but reside well within the domain of existence.
Now this part of the boundary is formed by a new type of black hole solutions.
These are counterrotating in the sense, that the horizon angular
velocity and the global angular momentum carry opposite signs
\cite{Kunz:2005ei}.
In Fig.~\ref{DE2} all counterrotating black holes
are represented by the shaded area.

\begin{figure}[h]
\begin{minipage}{17pc}
\includegraphics[width=13pc,angle=270]{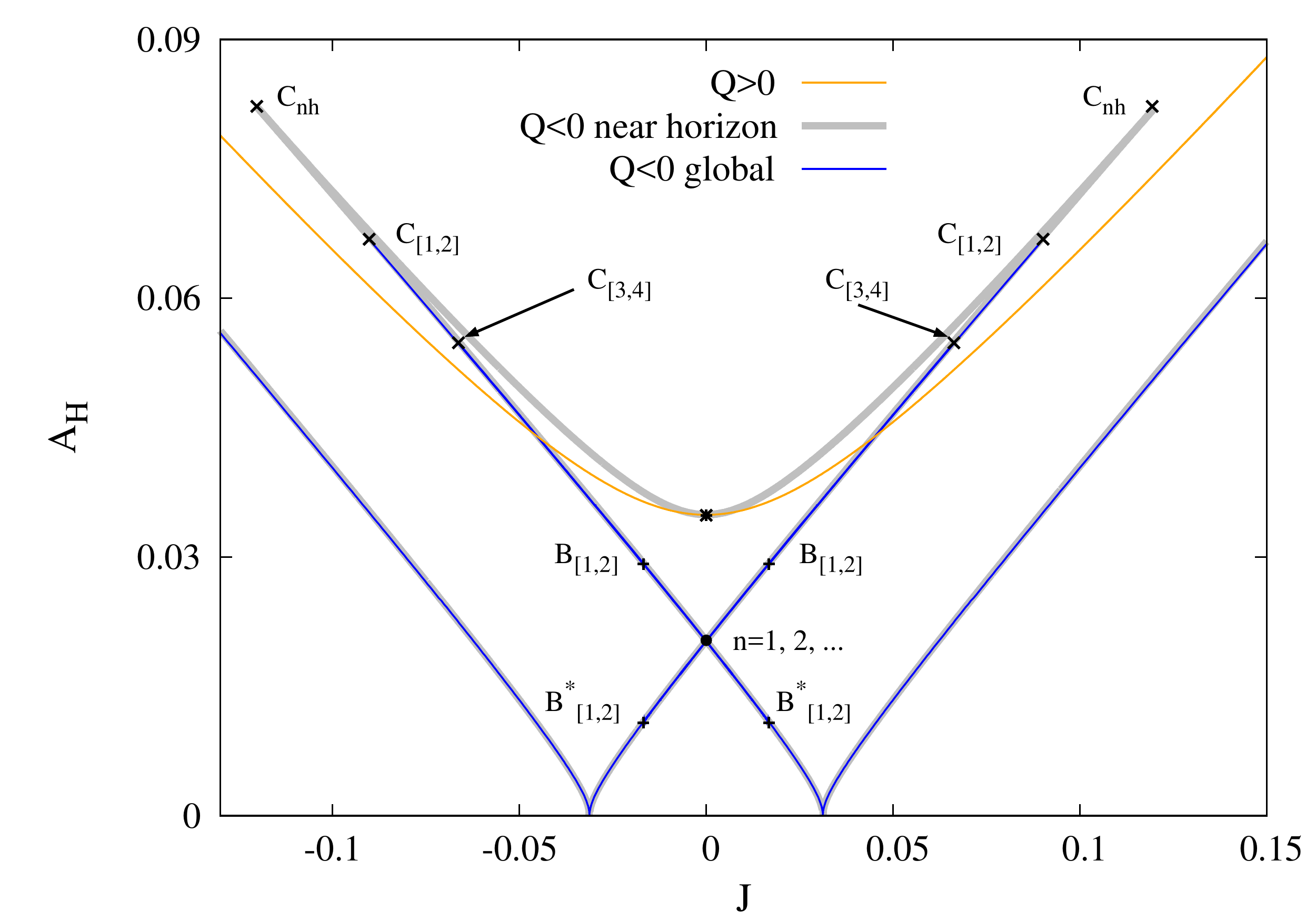}
\caption{\label{NH3}
$D=5$ EMCS near-horizon and global solutions:
area vs angular momentum ($\lambda=5$).
The asterisk marks the extremal static RN black hole.
}
\end{minipage}
\hspace{4pc}%
\begin{minipage}{17pc}
\hspace{-2pc}
\includegraphics[width=13pc,angle=270]{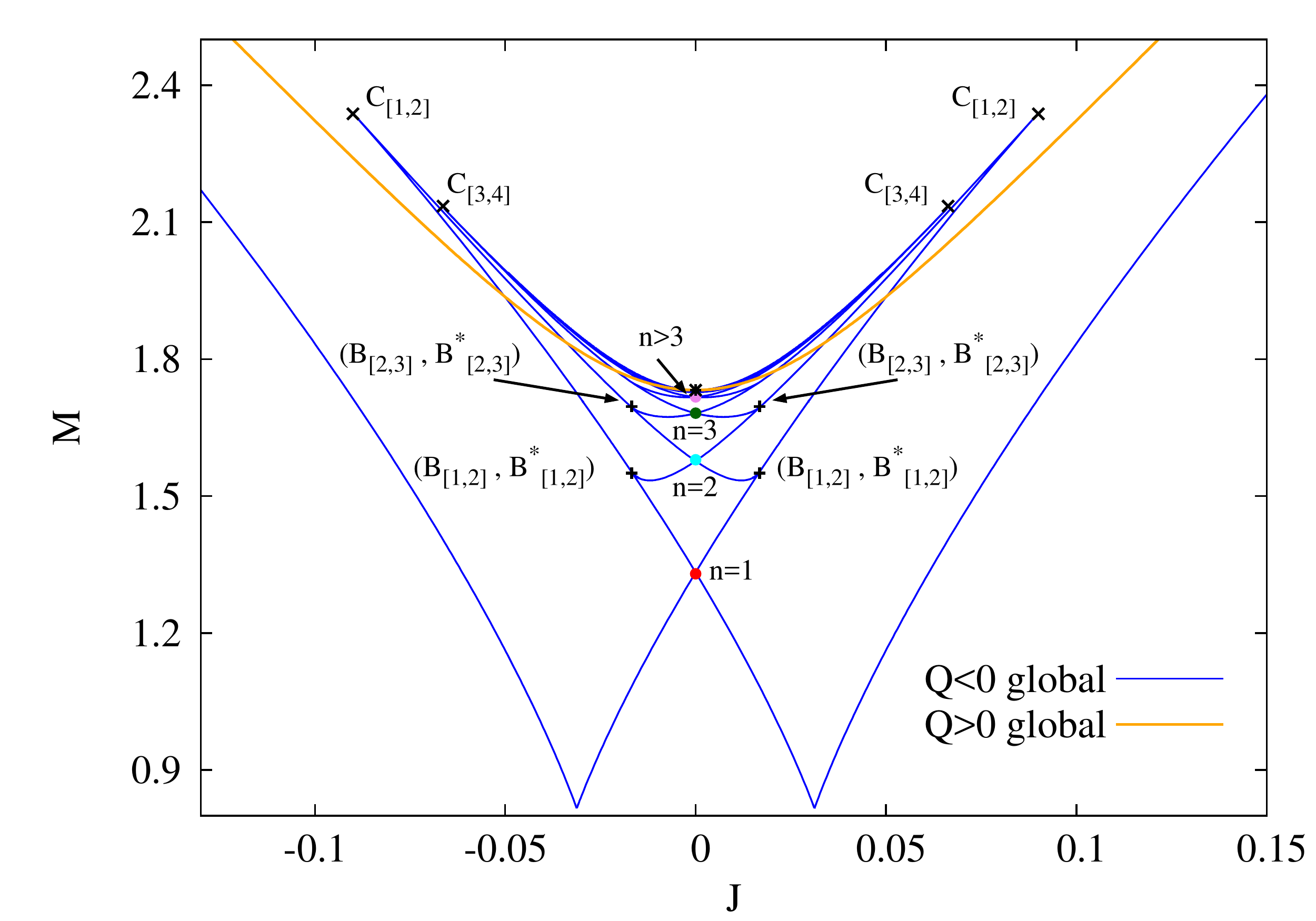}
\caption{\label{GS1}
$D=5$ EMCS global black hole solutions:
mass vs angular momentum ($\lambda=5$).
Bifurcation points $B$ and cusps $C$ are marked.
}
\end{minipage}
\end{figure}

For CS coupling $\lambda >2$ further surprises appear.
First of all, uniqueness of black holes with spherical
horizon topology no longer holds 
\cite{Kunz:2005ei}.
Here distinct black holes with the same global charges are present.
Second, there arises a strong mismatch
between the near-horizon solutions and the global solutions
\cite{Blazquez-Salcedo:2013muz,Blazquez-Salcedo:2015kja}.
In fact, there are near-horizon solutions that correspond to 
(i) no global solution, (ii) one global solution,
and (iii) many global solutions (possibly even
infinitely many).

The area versus the angular momentum
of the near-horizon solutions is illustrated in Fig.~\ref{NH3}
for CS coupling $\lambda=5$ for positive and negative charge.
For comparison also the global solutions are included in the figure.
Furthermore cusps and bifuraction points are noted.
In particular, while well connected with the branches
of near-horizon solutions, the RN solution is isolated
from the $Q<0$ global solutions.
Moreover, the $Q<0$ near-horizon solutions exhibit a 
non-static $J=0$ solution, which corresponds to a set of
distinct non-static global $J=0$ solutions, that can be
labeled by an integer $n$.
Thus there are black holes with a rotating horizon but
vanishing global angular momentum.
Also, the $Q<0$ solutions
always contain a degenerate ($J\to -J$) zero area solution.

The mass versus the angular momentum of the corresponding global solutions
is exhibited in Fig.~\ref{GS1}.
Here the degeneracy present in the area versus angular momentum
plot is lifted.
Instead an intricate branch structure of the global solutions is revealed.
In fact, in a regular manner, new branches of solutions arise,
which form cusps and then bifurcate with previous branches.
The cusps $C$ and bifurcation points $B$ are 
counted and labeled by integers in the figure.
When the branches cross, a degenerate pair of $J=0$ solutions appears,
also labeled by a repective integer $n$.
As $n$ increases, the mass of the rotating $J=0$ solutions tends to
the mass of the extremal RN black hole.

\begin{figure}[h]
\begin{minipage}{17pc}
\hspace{-1pc}
\includegraphics[width=14pc,angle=270]{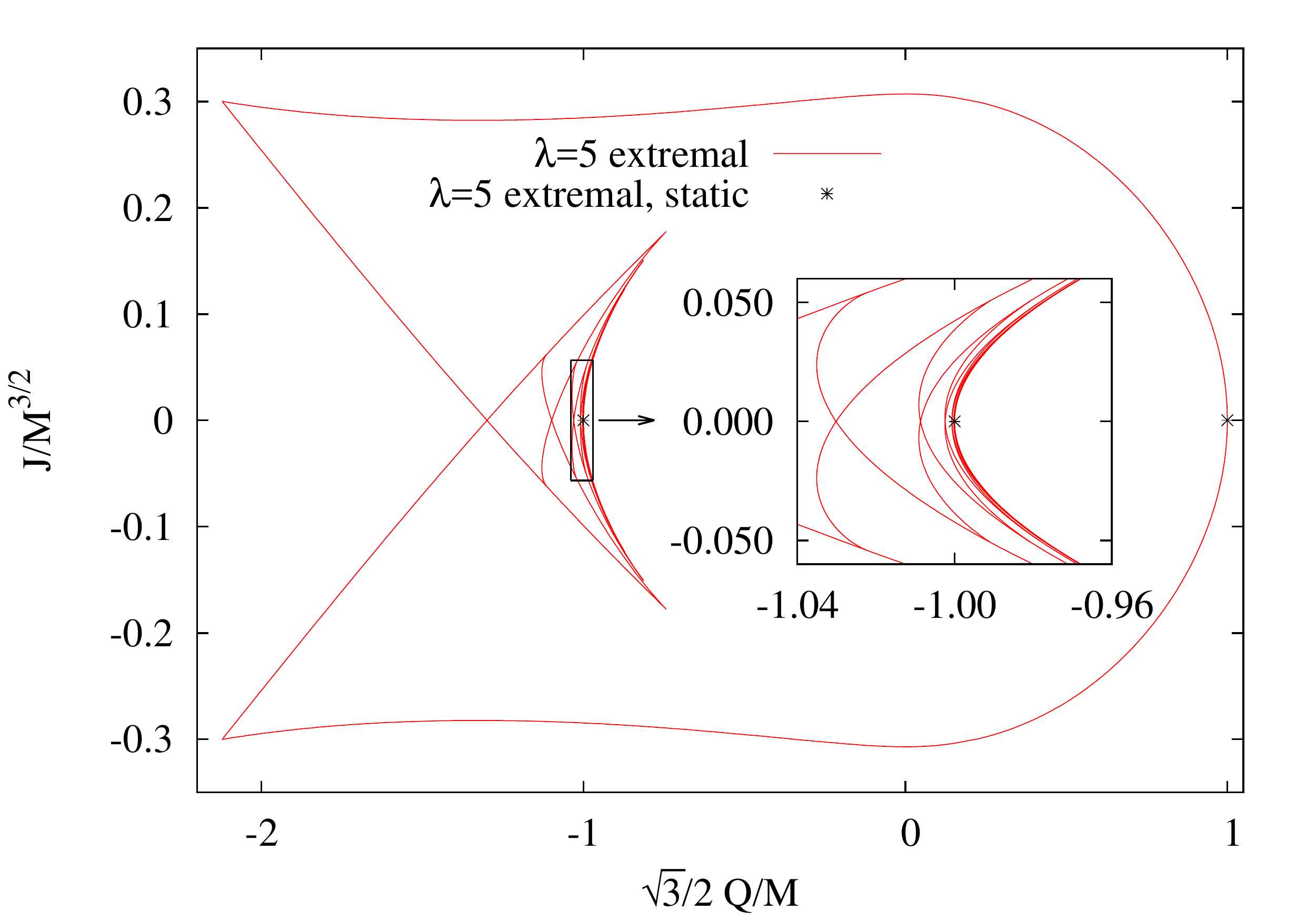}
\caption{\label{DE3}
Domain of existence of $D=5$ EMCS black holes:
angular momentum vs charge (scaled with appropriate powers of the mass)
for CS coupling $\lambda=5$.
}
\end{minipage}
\hspace{4pc}%
\begin{minipage}{17pc}
\hspace{-1pc}
\includegraphics[width=14pc,angle=270]{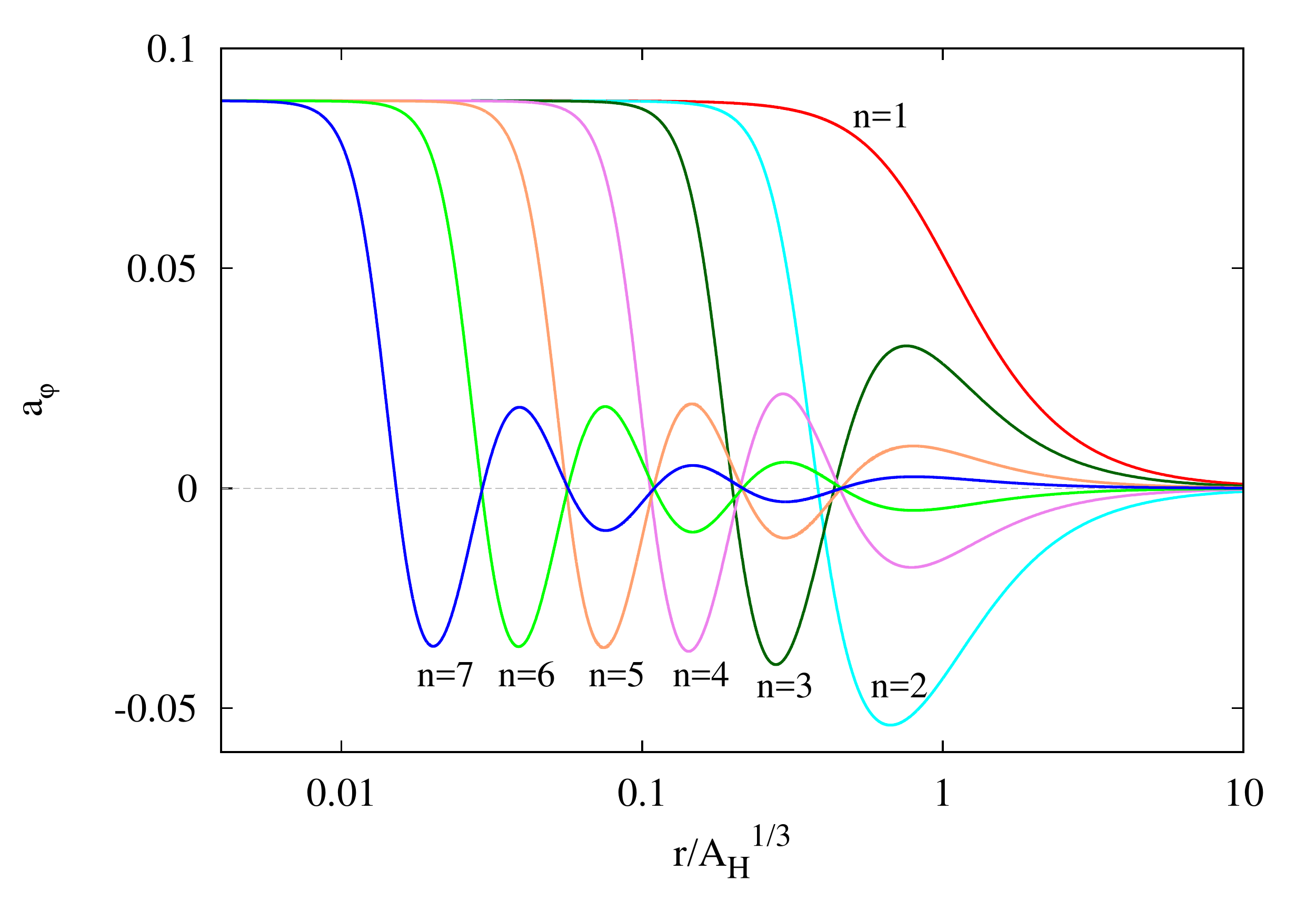}
\caption{\label{Nodes}
$D=5$ EMCS black holes:
Node structure of the gauge field function $a_\varphi$
($\lambda=5$, $n=1,\dots,7$). The numbering includes the node at
infinity.
}
\end{minipage}
\end{figure}

This intriguing branch structure for $Q<0$ is also seen in Fig.~\ref{DE3},
where the domain of existence of the EMCS solutions for 
$\lambda=5$ is shown.
While forming the boundary of the domain of existence,
extremal EMCS solutions reside also deep within this domain
along with the extremal static RN solution (black cross).
Only the extremal $n=1$ rotating $J=0$ solutions are part of the boundary.
The solutions with vanishing area represent the boundary solutions
at the cusps at maximal $Q/M$. 

When analyzing the branch structure one realizes, that the integer number
$n$ labeling the branches corresponds to the number of nodes
of the gauge field function $a_\varphi$ as illustrated in 
Fig.~\ref{Nodes} for $n=1,\dots,7$. While solutions with more than
thirty nodes have been constructed in a systematic way, one is lead to
conjecture, that there is in fact an infinity sequence of solutions,
$n=1,\dots,\infty$. Note, that the drag function of the metric
increases its nodes in an analogous manner.
This sequence of radially excited rotating black holes
is reminiscent of other physical systems with radial excitations,
such as the hydrogen atom.

\section{EMCS Solutions with AdS Asymptotics}

Let us now include a negative cosmological constant, $\Lambda = - 6/L^2$,
with AdS length scale $L$.
Then the solutions are no longer asymptotically flat but asymptotically
AdS.
The dS/AdS generalizations of the MP black holes are known in closed form
\cite{Hawking:1998kw,Gibbons:2004uw}.
But the corresponding EMCS black holes could be obtained in closed form
so far only for the case of gauged supergravity ($\lambda=1$),
where in a particular limit
they are known to preserve some amount of supersymmetry 
\cite{Gutowski:2004ez,Cvetic:2004hs,Chong:2005hr}.

\subsection{Charged solutions}

We first address the question of how the presence of the negative
cosmological constant affects the properties of the rotating
black holes discussed above.
The interesting new features present for the asymptotically flat solutions
are basically retained in the presence of the cosmological term.
In particular, when the CS coupling is sufficiently large,
there appear counterrotating black holes.
The black holes are no longer uniquely specified by their global charges.
Instead an analogous branch structure arises, where the solution
branches can be labeled by the node number of the 
magnetic gauge potential function.

\begin{figure}[h]
\begin{minipage}{38pc}
\hspace{-1pc}
\includegraphics[width=38pc,angle=0]{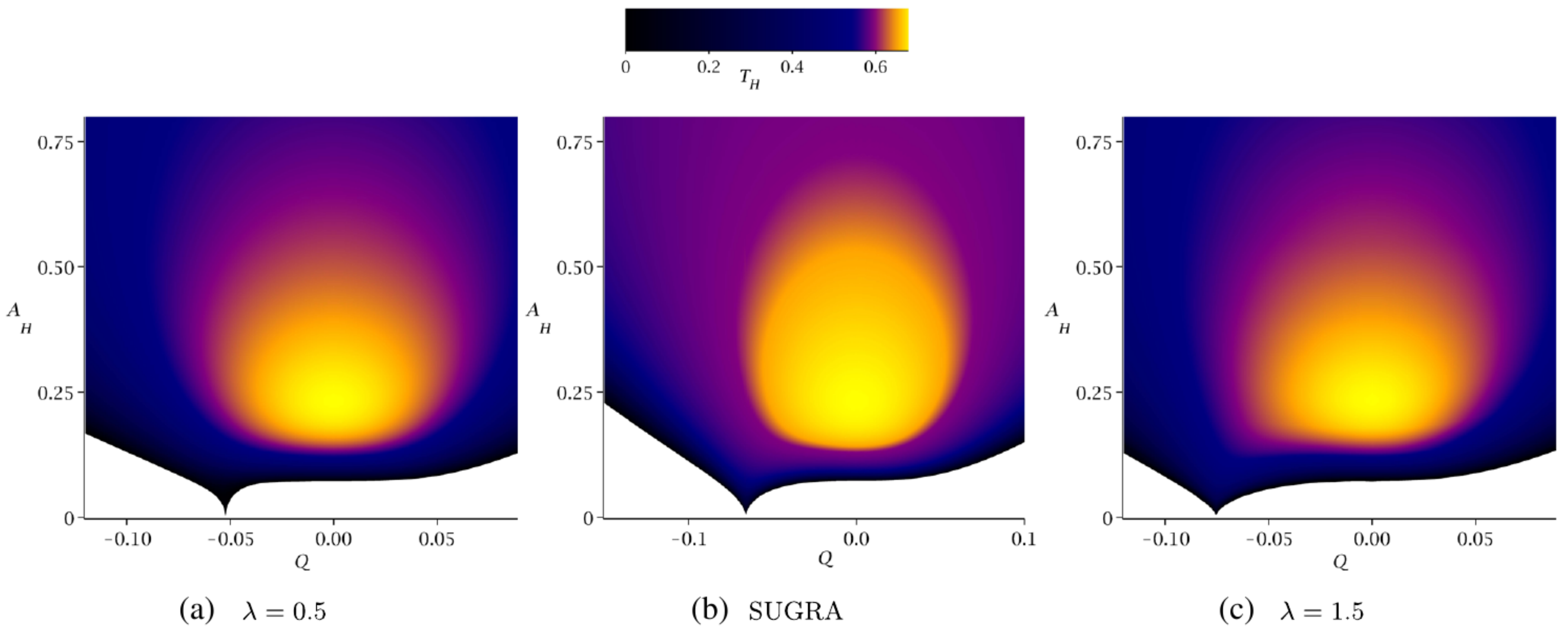}
\caption{\label{Lambda}
$D=5$ EMCS AdS black holes:
horizon area vs charge at different temperatures 
($J=0.00296$, $L=1$) for $\lambda = 0.5$ (a), $\lambda = 1$ (b), 
and $\lambda = 1.5$ (c).
}
\end{minipage}
\end{figure}

For CS coupling in the vicinity of the supergravity value, we demonstrate
in Fig.~\ref{Lambda} as an example
the $\lambda$-dependence of the area vs the charge and temperature.
The extremal solutions form the lower boundary ($T_H=0$).
The charge symmetry breaking by the CS term is again clearly visible,
with the zero area solutions only present for negative $Q$.

A new feature arises for small CS coupling $\lambda$.
Here a near-horizon analysis shows, that the two types of branches
(RN and MP for $\Lambda=0$) no longer cross. 
This leads to a gap in the set of regular extremal black hole solutions,
where the set of non-extremal solutions is limited by the so-called
gap set, which seems to consist of singular solutions
(as based on an extrapolation of the properties of the non-extremal solutions).

\subsection{Magnetized solutions}

Let us now include a new physical feature, namely a
new parameter $c_m$ of the AdS solutions associated
with the (finite) magnitude of the magnetic potential at infinity,
i.e., for $r \to \infty$: $a_\varphi \to c_m$.
This parameter therefore determines the magnetic flux through the
base space $S^2$ of the $S^1$ fibration of the $S^3$
(see Eqs.~\ref{at0}-\ref{at1})
\begin{equation}
\Phi_m = \frac{1}{4\pi} \int_{S^2_\infty} {\cal F} = -\frac{1}{2} 
c_m .
\end{equation}

For simplicity we first address the case of static EM solutions,
since, acting like a box,
the AdS background also allows for static solitonic EM solutions
\cite{Herdeiro:2015vaa}.
In particular, by considering magnetic fields only,
one obtains regular EM solitons in $D=5$ (and higher odd dimension),
which are characterized by the magnetic flux parameter $c_m$
\cite{Blazquez-Salcedo:2016vwa}.
Interestingly, there is a maximal value of $c_m$, where such solitons
exist.

\begin{figure}[h]
\begin{minipage}{17pc}
\hspace{-1pc}
\includegraphics[width=18pc,angle=0]{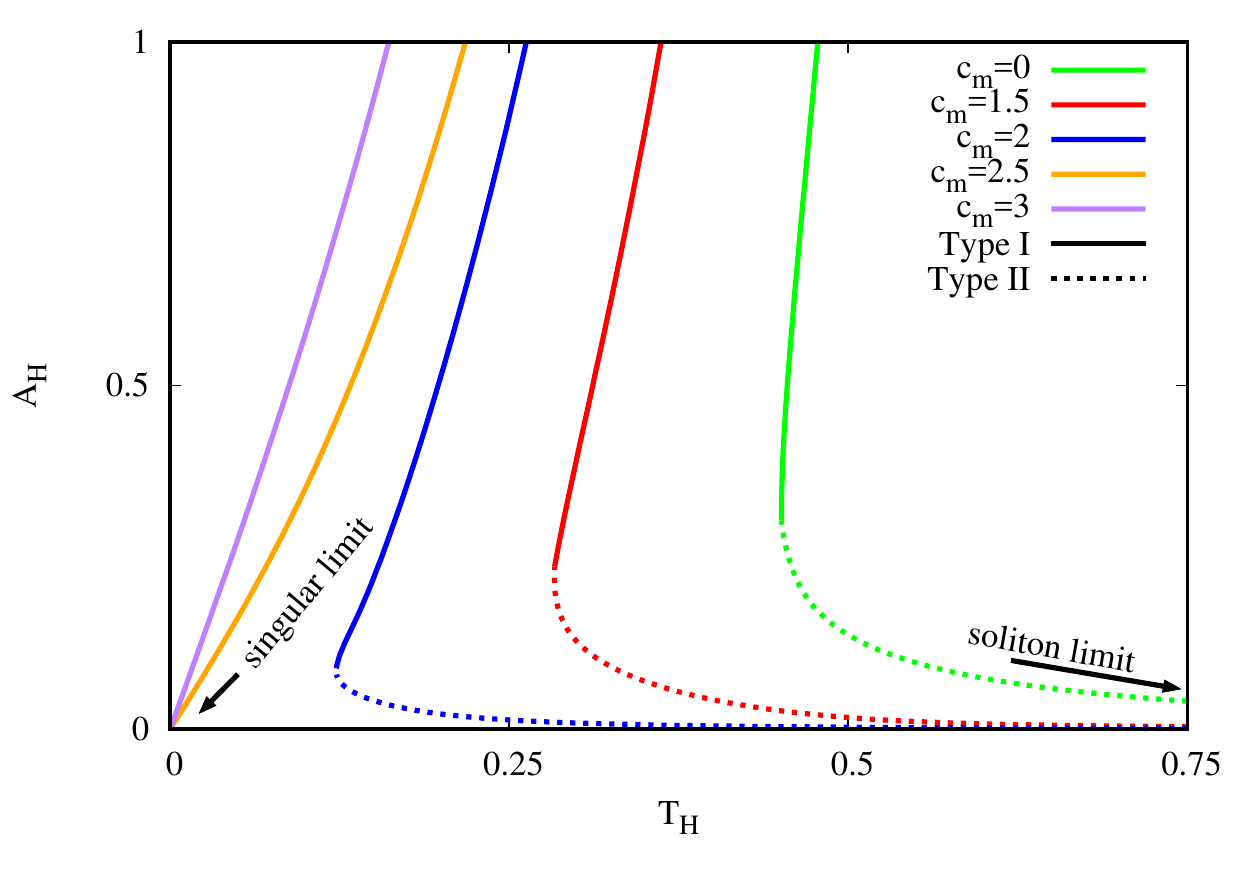}
\caption{\label{areal}
Static magnetized AdS EM black holes:
area vs temperature (magnetic flux parameter $c_m=0$ to 3).
}
\end{minipage}
\hspace{4pc}%
\begin{minipage}{17pc}
\hspace{-1pc}
\includegraphics[width=18pc,angle=0]{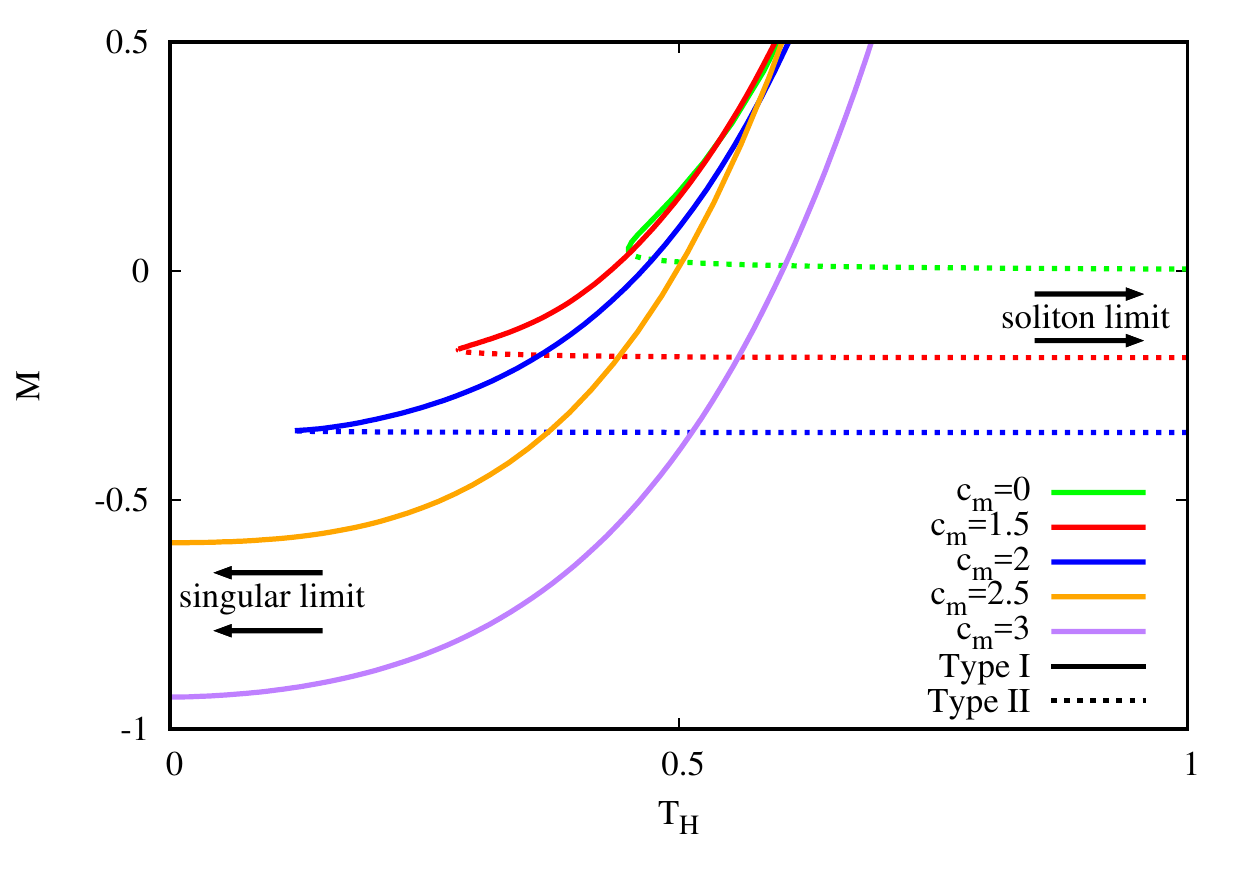}
\caption{\label{massl}
Same as Fig.~\ref{areal} for the mass vs the temperature.
Singular and soliton limits are indicated.
}
\end{minipage}
\end{figure}

By imposing a regular horizon, static EM black hole solutions arise,
which can be classified into two types. 
Figs.~\ref{areal} and \ref{massl} show
their horizon area and mass versus their temperature.
In type I black holes the horizon area and the mass
increase with the temperature. 
In contrast, in type II black holes both decrease
as the temperature increases. 
Type II black holes can be deformed
continuously into solitons, when the horizon size approaches zero.
Thus they exist only below the maximal value of $c_m$ for solitons.
Type I black holes on the other hand exist also for large values of $c_m$,
but they become singular in the limit of vanishing horizon size.

Clearly the above black holes represent new families of static EM AdS black holes,
whose properties differ from those of the known RN AdS black holes.
This suggests that similar magnetized EMCS AdS solutions could also exist.
However, in the EMCS case solutions with a magnetic field
must necessarily rotate. 
Indeed rotating EMCS generalizations of the above 
soliton and black hole solutions exist
\cite{Blazquez-Salcedo:2017cqm}.
These solutions then also carry an electric charge,
where we should now distinguish between the ordinary charge,
the Page charge, and the $R$ charge
\begin{equation}
Q^{(R)}= 
- \frac{1}{2} \int_{S_{\infty}^{3}} \left( \tilde {\cal F }
+\frac{2\lambda}{3\sqrt{3}} {\cal A} \wedge {\cal F} \right) ,
\end{equation}
since the second term now contributes, and its
prefactor differs for these three types of charges.

\subsection{Magnetized squashed solutions}

Let us now consider a last twist w.r.t.~the plethora of EMCS solutions
by considering solutions which are asymptotically only locally AdS.
Thus we impose that the magnetized
rotating EMCS solutions should asymptotically approach not a round $S^3$ sphere
but instead a squashed sphere.
As discussed in \cite{Cassani:2014zwa}
the boundary metric can then be expressed as
\begin{equation}
ds^2_B = L^2d\Omega_{(v)}^2- dt^2,~~~~d\Omega_{(v)}^2=
\frac{1}{4}\left( d\theta^2+\sin^2\theta d\phi^2 
+ (d  {\psi} + v \cos\theta d \phi)^2 \right) ,
\end{equation}
where $v$ is a control parameter.
Clearly, for $v=1$ the $S^3$ sphere becomes round,
whereas for $v=0$ the solutions tend to AdS black strings
and vortices, their boundary corresponding to $S^2 \times S^1$.

Then for a given $\lambda$ a family of squashed magnetized soliton solutions 
arises smoothly from the AdS vacuum, 
which are labeled by the parameter $v$ and satisfy 
the relation $J= \Phi_m Q^{(R)}$.
In the case of gauged supergravity ($\lambda=1$)
these solitons retain some amount of supersymmetry
\cite{Cassani:2014zwa}.
In fact, the properties of these squashed susy solitons can be 
expressed simply in terms of the squashing parameter $v$.

\begin{figure}[h]
\begin{minipage}{17pc}
\hspace{-1pc}
\includegraphics[width=14pc,angle=270]{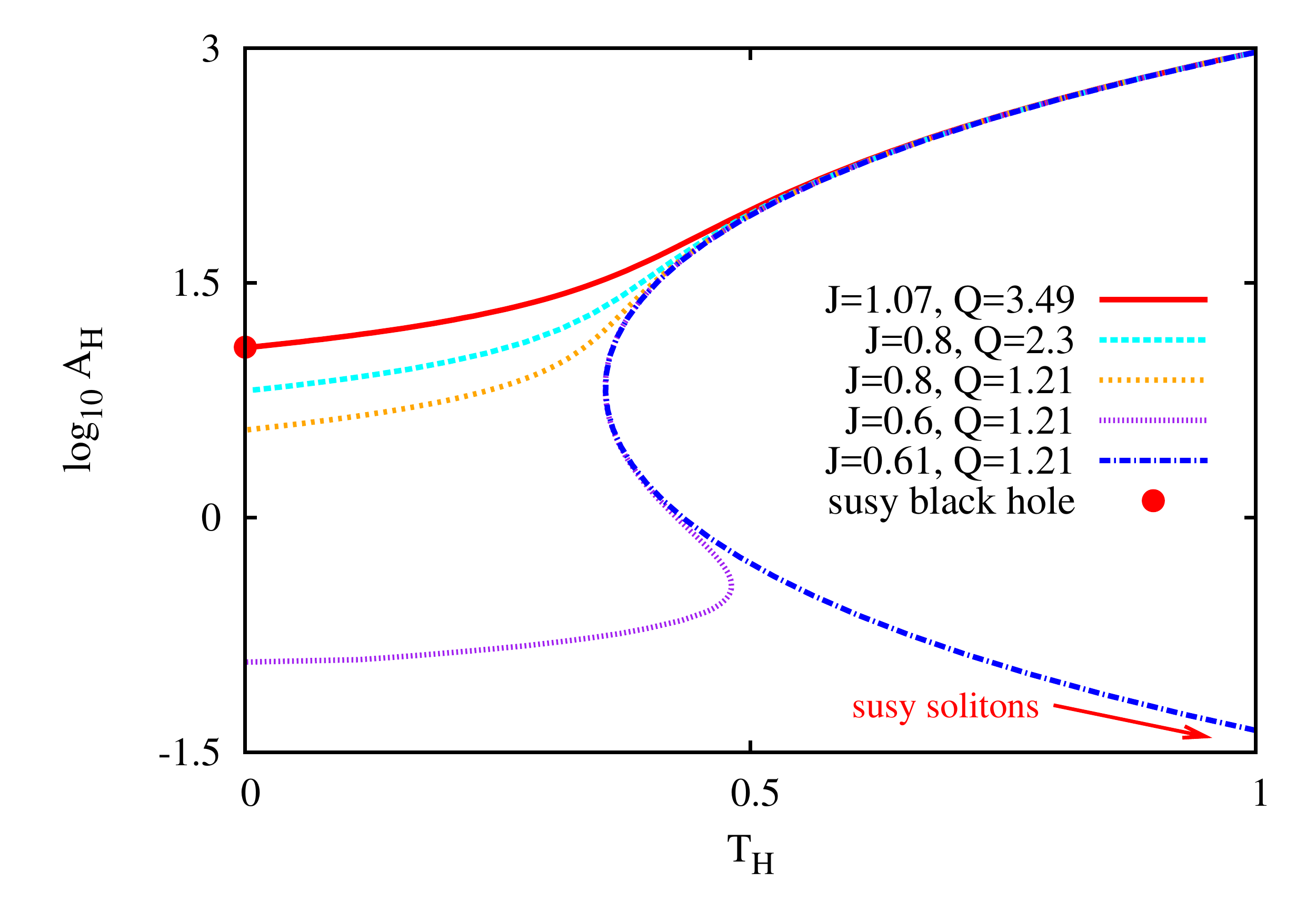}
\caption{\label{areals}
Squashed magnetized EMCS AdS black hole solutions:
area vs temperature ($\lambda=1$, $L=1$, $v=1.65$, $c_m=-1$).
}
\end{minipage}
\hspace{4pc}%
\begin{minipage}{17pc}
\hspace{-1pc}
\includegraphics[width=14pc,angle=270]{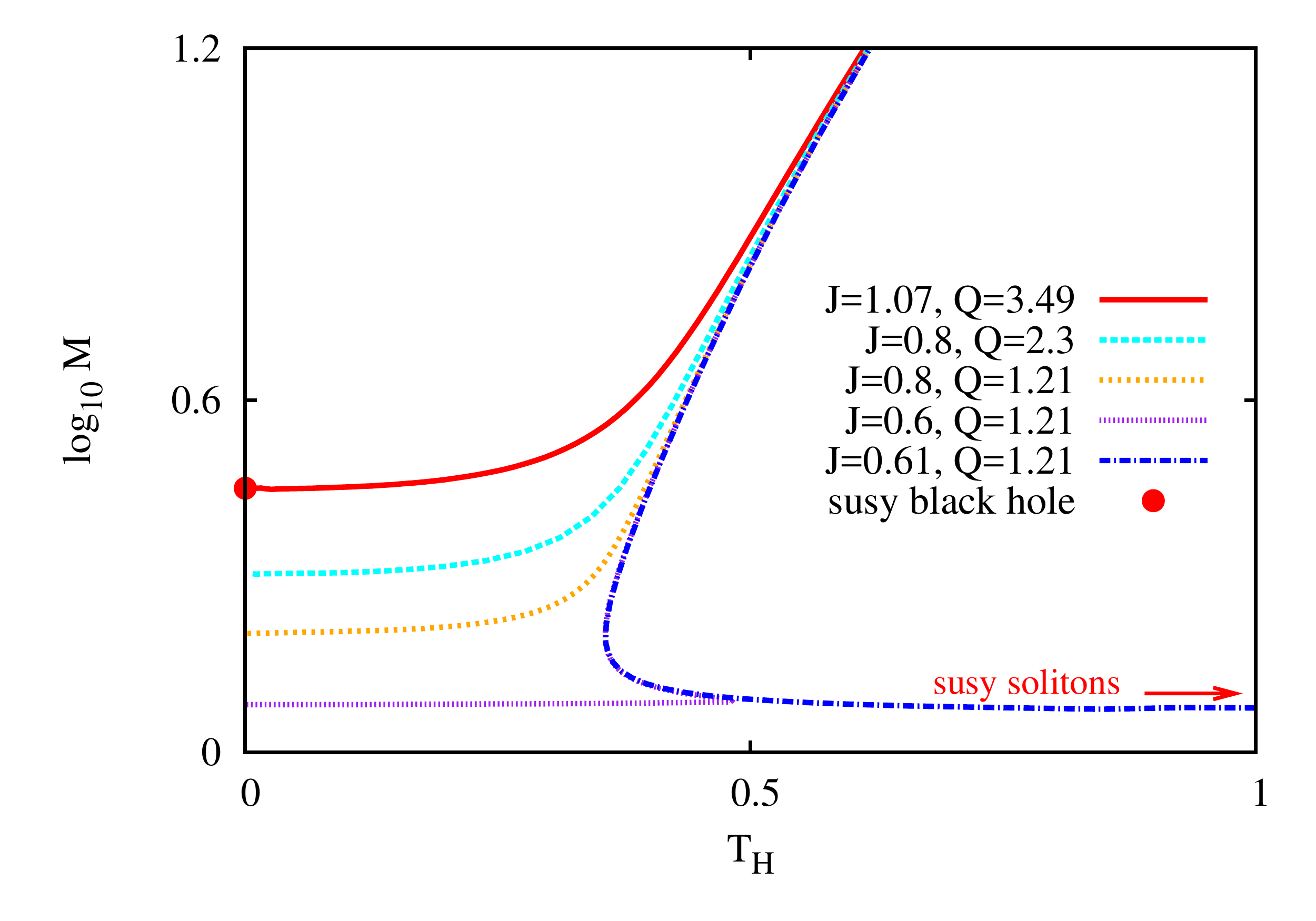}
\caption{\label{massls}
Same as Fig.~\ref{areals} for the mass vs the temperature.
The red dot marks a susy black hole.
}
\end{minipage}
\end{figure}

As one might expect, the solitons are related to a new class of squashed
magnetized black holes. In the general case, these black holes
are characterized by their mass, charge, angular momentum and magnetic
flux parameter. In the limit of vanishing horizon radius 
solitons can be approached for an appropriate parameter choice.
Families of black hole solutions for the case of gauged supergravity
are exhibited in Figs.~\ref{areals} and \ref{massls}.

Generically the black holes possess an extremal limit with a finite
horizon area. Among these extremal black holes a particular family
stands out: the supersymmetric black holes.
For these solutions the trace of the boundary stress tensors
vanishes, yielding the condition $c_m= \pm \frac{L}{\sqrt{3}} (1-v^2)$.
Like the susy solitions these susy black holes can be characterized
by the squashing parameter $v$.

This observation then leads to the following new picture for the 
classes of supersymmetric EMCS AdS black holes.
Besides the previously known family of Gutowski-Reall black holes
there exists a new family of squashed magnetized supersymmetric black holes.
In fact, the new susy squashed magnetized black holes intersect the
Gutowski-Reall black holes at a critical configuration, where 
the squashing and magnetization vanish, i.e., $v=1$ and $c_m=0$.

Obviously, there are still many unanswered questions, and many avenues
are open awaiting further investigations.
This holds, in particular, w.r.t.~the relevance 
of the above configurations in an AdS/CFT context.


\subsection*{Acknowledgments}
We gratefully acknowledge support by
the DFG Research Training Group 1620 ``Models of Gravity''.
 E. R. acknowledges funding from the FCT-IF programme. F. N.-L. acknowledges funding from Complutense University - Santander under
project PR26/16-20312.

\end{document}